\documentclass[aps,prl,showpacs,twocolumn,reprint,floats,epsfig,pdflatex,superscriptaddress]{revtex4-1}
\usepackage[]{csquotes}

\usepackage{epsfig}
\usepackage{amsmath}
\usepackage{amsfonts}
\usepackage{graphicx}
\usepackage{amssymb}
\usepackage{amsbsy}
\usepackage{subfigure}
\usepackage{tabularx}

\begin{document}

\title {Absolute Te$_2$ reference for barium ion at $455.4~$nm}

\author{T.~Dutta}
\author{D.~De~Munshi} 
\affiliation{Centre for Quantum Technologies, National
University Singapore, Singapore 117543}
\author{M.~Mukherjee}
\affiliation{Centre for Quantum Technologies, National University Singapore, Singapore 117543}
\affiliation{Department of Physics, National University Singapore, Singapore 117551. }
\affiliation{MajuLab, CNRS-UNS-NUS-NTU International Joint Research Unit, UMI 3654, Singapore}

\date{\today}
%
%
%


%
%
%
%
%

\begin{abstract}
Precision atomic spectroscopy is presently the work horse in quantum information technology, metrology, trace analysis and even for fundamental tests in physics. Stable lasers are inherent part of precision spectroscopy which in turn requires absolute wavelength markers suitably placed corresponding to the atomic species being probed. Here we present, new lines of tellurium (Te$_2$) which allows locking of external cavity diode laser (ECDL) for precision spectroscopy of singly charged barium ions. In addition, we have developed an ECDL with over 100 GHz mod-hop-free tuning range using commercially available diode from $\textit{Nichia}$. These two developments allow nearly drift-free operation of a barium ion trap set-up with one single reference cell thereby reducing the complexity of the experiment.
\end{abstract}



\maketitle


\section{Introduction}

Ion trap precision spectroscopy has led the way to implement quantum algorithms~\cite{Blatt:08}, test fundamental symmetries of nature~\cite{Lean:11}, trace element analysis~\cite{Trac:12}, isotope separation~\cite{list:03} \textit{etc.}. In each of these experiments an essential component has been a stable laser to probe atomic or molecular transition. Stability of a laser is judged by its emission bandwidth as well as slow drift of its wavelength. The emission bandwidth is narrowed by locking to high finesse optical cavity which can currently achieve sub-Hz linewidth in short time scales~\cite{clock:13}. However the emission wavelength of the laser locked to a cavity can drift due to ambient temperature fluctuation, low frequency mechanical vibrations \textit{etc.}. There has been tremendous development in building ultra-stable reference cavities which can restrict these drifts below kHz/day~\cite{clock:13}. An alternative method to restrict the drift is to actively lock the laser to known frequency reference of an atom or molecule. Particularly for experiments which require extended period of data aquisition, active locking to atomic or molecular reference is preferred due to their robustness. Furthermore to avoid complexity in the experimental setup it is preferred to have the same atomic/molecular reference cell for all involved transitions. However, this is not always possible due to lack of suitable transitions in a single atom or molecule. Mostly iodine dimers and tellurium dimers are used as choice of reference apart from hollow cathode lamps for different elements. The later requires opto-galvanic detection setup while the former relies on Doppler free optical spectroscopy. In the following we have implemented modulation transfer spectroscopy~(MTS) which unlike the frequency modulation spectra~(FMS) produces a zero crossing signal at the resonance frequency thereby allowing direct frequency locking of the laser to the molecular transition frequency similar to a Pound-Drever-Hall signal~\cite{pdh:83}. \\
Tellurium dimer has a rich spectra covering parts of ultra-violate~(UV) and visible wavelengths. A comprehensive study of its broad spectra has been performed by Cariou and Luc in what is now known as the Te$_2$ atlas \cite{Te2atlas:80}. However, in order to frequency lock a laser it is important to detect transition lines close to the targeted transition line of the atomic species under investigation. Russell J. De Young has performed absorption spectroscopy above $500~$nm to extract the absorption cross-section to the first electronic excited state~\cite{You:94}. T.~J.~Scholl \textit{et al.} measured $39$ lines between $420~$nm to $460~$nm to cover the Stilbene-420 dye tuning curve employing saturation spectroscopy~\cite{Sch:05}. Tellurium in addition to the thorium and uranium emission atlas of Los Alamos provide suitable reference below $500$~nm which includes Gillaspy and Sansonetti's measurement between between $471-502~$nm for Coumarine dye at $480~$nm~\cite{Gil:91} and Courteille \textit{et al.}'s measurement close to $476~$nm for diode laser locking for $Yb^+$ ion spectroscopy~\cite{Ma:93}. In the region of interest for hydrogenic atoms like deuterium, hydrogen and positronium ranging from $486~$nm to $488~$nm a number of experiments has been performed~\cite{Mct:90}. C.~Raab~\textit{et al.} performed precision measurement on Te$_2$ spectra close to the Ba$^+$ ion Doppler cooling transition at $493~$nm~\cite{Raa:98}. More recently, J.~Cooker~\textit{et al.} showed a commercial blue laser diode diode stability locked to Te$_2$ line at $444.4~$nm for diode laser reference in transfer cavities~\cite{Coo:11}. In the meanwhile I.~S. Burns~\textit{et al.} extended the Te2 spectral reference close to $410~$nm where commercial blue diodes are now available~\cite{Bur:06}. \\
In this work we extend the available Ba$^+$ spectroscopic tool further by adding new Te$_2$ spectral lines close to the $S_{1/2}-P_{3/2}$ transition in barium ion at $455.4~$nm. In order to drive this transition, we have developed an extended cavity diode laser employing commercially available violet laser diode at $455~$nm with mode-hope-free turning range of more than $100~$GHz. In order to determine the absolute wavelength, simultaneous opto-galvanometric measurement of a barium hollow cathode lamp~(HCL) was recorded. We find two new Te$_2$ lines within the $1.2~$GHz wide HCL spectrum with the closest one being only $79~$MHz away from the needed barium line. The closeness of this transition makes it a suitable frequency reference which can easily be bridged by an acousto-optic modulator~(AOM). In the following, we will provide a description of our setup, measurement procedure and present our results before concluding in the last section.

\section{Experimental setup}
\label{sec:setup}

\begin{figure}[htbp]
\centering
\fbox{\includegraphics[width=\linewidth]{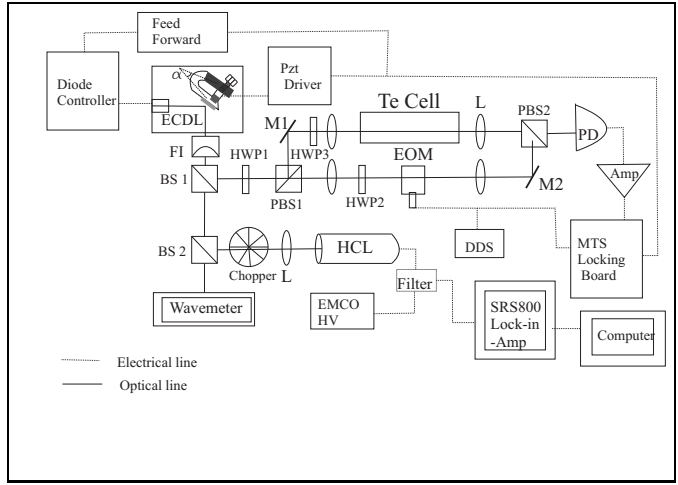}}
\caption{layout of Experimental set up;ECDL-external cavity diode
laser,FI-Faraday isolator,PBS-polarising beam splitter,BS-beam
splitter,HWP- half wave plate,M-Mirror,L-lens,EOM-electro optic
modulator.} \label{fig:1}
\end{figure}

\begin{figure}[htbp]
\centering
\fbox{\includegraphics[width=\linewidth]{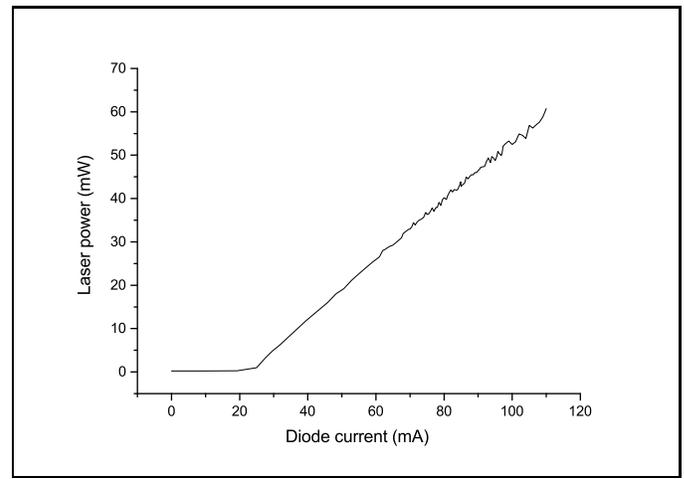}}
\caption{ECDL output: The output power of the ECDL as a function of the input current. The small jumps in the output power observed above $50~$mA of drive current indicate mode-hops in the ECDL cavity. similar mode-hops due to grating angle are suppressed by feed-forward applied to the current.}
\label{fig:2}
\end{figure}

In this section we briefly describe our laser design and electronics to generate the $455.4~$nm laser light along with an overview of the setup to measure the MTS spectra of Te$_2$ in a hot vapor cell. A schematic of the experimental setup is shown in fig.~\ref{fig:1} which includes both the MTS setup of tellurium and the HCL spectroscopy of barium using the $455~$nm diode laser. The laser is an extended cavity diode laser similar to the NIST design~\cite{Wie:91} where a special pivot point is selected to have minimal cavity length change as the ECDL is frequency scanned. The laser diode is \textit{Nichia NDB4216E} with anti-reflection coating on the front facet in order to minimize the diode laser modes. The diode is driven by an in-house $CQT$ designed current drive, very similar to the original J.~Hall design~\cite{Lib:93} while mode-hop-free operation over a wide frequency range is ensured by feed forward added to the diode operating current. The feed forward current and the scan voltages are generated from a direct digital synthesis~(DDS) implemented on low cost $Arduino-UNO$ board taking into account the ECDL cavity length change as a result of the angular ($\alpha$) tuning of the piezo. The optical output power of the diode with and without the ECDL is shown in figure~\ref{fig:2}. It is clear that the laser diode under ECDL condition shows more regular mode-hops as the current is increased beyond $60~$mA. At the operational wavelength, total power of $50~$mW is available at a diode current of $100~$mA for the experiment. Out of this, about $10~$mW is used for the implementation of the MTS setup, about $3~$mW is used for the HCL spectrometry, about $15~$mW is used for wavemeter measurement and the rest is available for the ion trap experiment. All these paths are fiber coupled with an efficiency of about $40\%$. The unusually high power requirement in the wavemeter path is mainly due to low efficiency in the wavemeter switch which is located about $50~$m away from our laboratory.\\

A faraday isolator (FI) is placed in front of the ECDL minimizing the optical feedback. As shown in fig.~\ref{fig:1}, the laser beam from the ECDL is divided into three components by a couple of beam splitters(BS) BS1 and BS2 for the MTS setup, the HCL setup and the $30~$MHz resolution wavemeter setup respectively. The optical setup for MTS spectroscopy utilizes two polarising beam splitter (PBS): PBS1 is used to split the beam into pump and probe while PBS2 is used for re-combining them as both arms overlap inside the Te$_2$ cell. The intensity ratio in those two beams is controlled by a zero order half wave plate (HWP) HWP1. The other two half waveplates HWP2 and HWP3 are used to control the polarization of individual beams. The pump beam is phase modulated by an electro-optic modulator~(EOM)(crystal:Mgo doped LiNbO3) which is driven at a modulation frequency of $5.8~$MHz. The probe beam is aligned collinearly with the counter propagating modulated pump beam through a $10~$cm long Te$_2$ cell which is placed inside an oven heated to $530~$K. The temperature of the cell is maintained to within $0.5^\circ$K by thermal isolation. Two photodiodes of responsivity $0.64~$A/W detect the MTS signal or saturation absorption signal after the PBS2. The photodiode (PD) signal is then amplified by a low noise amplifier and fed into a $CQT-$built MTS Locking board comprising of a frequency mixer with low pass
filter at $30~$kHz cut-off and a PID controller for frequency locking purposes. This board generates error signal which is split into two parts: one feeds back to the current of the ECDL for any fast frequency correction and the other feeds back to the piezo driver board for slow drift corrections. The light which is reflected from BS2 is used for HCL spectroscopy. This part is sent to the barium hollow cathode lamp after chopping at a frequency of $1~$kHz to avoid low frequency electronic noise in the lock-in-amplifier detection setup. The opto-galvanic signal from the HCL is separated out using a high pass filter and the voltage drop across a $15~$kOhm resistor is detected by \textit{Stanford Research:SRS380} lock-in-amplifier. Simultaneous data of the lock-in-amplifier and the wavemeter are logged into a computer using python code.

\section{Measurement procedure and results}
\label{sec:measpro}

\begin{figure}[htbp]
\centering
\fbox{\includegraphics[width=\linewidth]{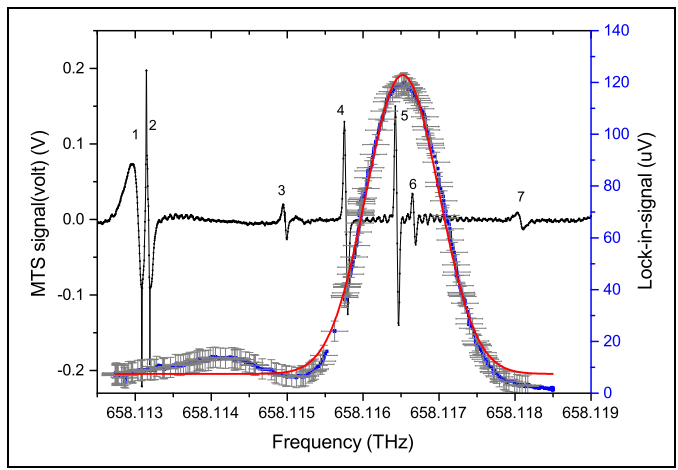}}
\caption{Spectra: The Te$_2$ MTS spectra shown as a function of laser frequency. The barium hollow cathode lamp spectra is also shown here as a reference for the frequency axis. Line no. 1 corresponds to the only known line of the Te$_2$ atlas. The line of interest is the line no. $5$ which is about $79~$MHz from the barium line center.}
\label{fig:3}
\end{figure}

\begin{figure}[htbp]
\centering
\fbox{\includegraphics[width=\linewidth]{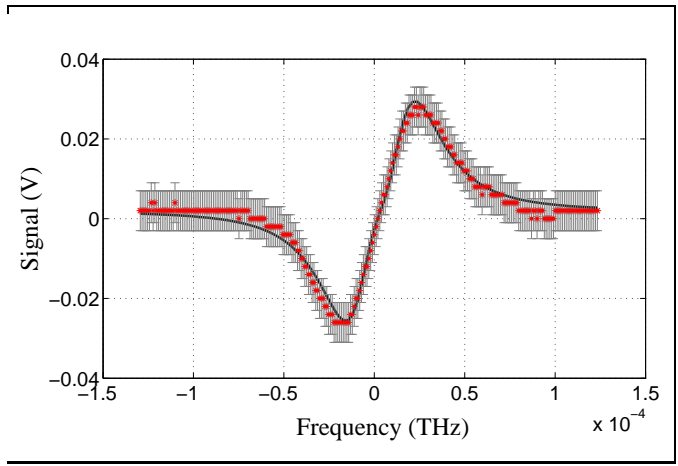}}
\caption{Spectra: The Te$_2$ MTS spectra along with the line-shape fit. The $*$ denotes experimental data along with errorbar while the solid line shows the fit with reduced $\chi^2$ of $0.98$. The obtained linewidth from the fit is $20.9(4)~$MHz.}
\label{fig:4}
\end{figure}

A symmetric triangular voltage $0-10~$V is applied to the piezo of the ECDL in order to scan the range of frequencies which is close the barium resonance as shown in Fig.~\ref{fig:3}. In order to obtain this full range of frequencies spanned over $7~$GHz, appropriate feed-forward has been applied. Both the tellurium resonances as well as the HCL spectra of the $S_{1/2}-P_{3/2}$ transition of the barium ion observed within this scan range. The frequency axis is calibrated by locking the laser at individual Te$_2$ resonance and measuring the wavelength from the wavemeter with a resolution of $30~$MHz. The absolute wavelength value has been determined from the Gaussian fit to the HCL spectra which matches well with NIST data. The linearity of the frequency axis is determined from the fit with reduced $\chi^2\approx0.98$ leading to $1\%$ uncertainty. As is evident from the HCl spectrum containing the Gaussian fit, the barium resonance wavelength can be determined within $20~$MHz uncertainty, thereby limiting our overall uncertainty of the absolute scale to the same value. In order to ensure that individual Te$_2$ lines are within the uncertainty set by the HCL spectrum, we performed a line-shape fit to Te2 line as shown in figure~\ref{fig:4}. The line shape of MTS resonance is devoid of any background slope unlike FMS. This shape can be described according to \cite{line:82} by

\begin{eqnarray}
S(\Delta)&=& Re\Bigg[ \sum_{j=a,b}\frac{\mu_{ab}^2}{\gamma_j+i\delta}\Bigg( \frac{1}{\gamma_{ab}+i(\Delta+\delta/2)} - \frac{1}{\gamma_{ab}+i(\Delta+\delta)} +\nonumber\\
&& \frac{1}{\gamma_{ab}-i(\Delta-\delta)} - \frac{1}{\gamma_{ab}-i(\Delta-\delta/2)} \Bigg) e^{-i\theta} \Bigg],
\label{eq:1}
\end{eqnarray}

where, $a$ and $b$ denotes the electronic levels of Te$_2$, $\mu_{ab}$ is the electric dipole matrix element between them, $\Delta$ is the laser detuning, $\gamma_{j}$ are decay rates of the levels and $\gamma_{ab}$ is the optical relaxation rate between $a$ and $b$. The modulation frequency and phase are given by $\delta$ and $\theta$ respectively. As an example one of the the resonance line shape as obtained in the experiment is shown in fig.~\ref{fig:4}. The line is fitted with a overall scaling factor equivalent to $\sum_{j=a,b}\frac{\mu_{ab}^2}{\gamma_j+i\delta}$, the relaxation rate $\gamma_{ab}$ for the involved transition and the unknown phase $\theta$. The relaxation rate $\gamma_{ab}$ obtained from the best fit provides the linewidth to be $20.9(4)~$MHz with a reduced $\chi^2\approx 0.98$. The zero-crossing of the resonance along with electronic suppression (about 100) allows the laser to be locked with a bandwidth of a few hundred kHz. The model mostly fits well with all the resonances except a few where an additional etalon effect modifies the base level of the signal. One particular point to note is the width of these resonances are higher as compared to the Te$_2$ resonances obtained near the barium S$_1/2$-P$_3/2$ transition at $493~$nm which is attributed to the higher vibrational level densities for shorter wavelengths.\\

\begin{table*}[htbp]
\centering
\caption{\bf Tellurium reference lines relative to barium resonance line: The line no. $1$ corresponds to line no. $1677$ in the tellurium atlas~\cite{Te2atlas:80}. All other lines are observed for the first time. Line no. $5$ is closest to the barium transition where spectroscopic laser is frequency locked. The first two lines here are unresolved due to fast scan rate and their relative strength are given as "sat" meaning saturated with respect to line no. $5$.}
\begin{tabular}{cccc}
\hline
line number & Relative frequency/GHz & Wavenumber/cm$^-$ & Relative strength \\
\hline
 1 & 3.423 & 21952.29007 &  sat \\
 2 & 3.341 & 21952.29281 &  sat \\
 3 & 1.542 & 21952.35282 & 1.6 \\
 4 & 0.749 & 21952.37927 & 8.8 \\
 5 & 0.079 & 21952.40162 & 10 \\
 6 & - 0.149 & 21952.40922 & 2.3 \\
 7 & - 1.548 & 21952.45589 & 0.8 \\
\hline
\end{tabular}
  \label{tab1}
\end{table*}

The result as summarised in figure~\ref{fig:3} contains seven resonances which has been observed for the first time. The previously reported lines in the Te$_2$ atlas are about $3.5~$GHz away from the barium line, therefore it cannot be used for laser locking. However for the linearity check of our frequency scan we have also used those resonance and found them to be matching well within the uncertainties. The frequencies are tabulated in table~\ref{tab1} along with their uncertainties and relative strengths. As for line no. $1$ and $2$, the resonance lines are close to each other due to our fast scan. We have observed these lines with higher resolution scan as well. Notably, despite the high relative strength of the second line, it was not observed in the $Te_2$ atlas possibly due to poorer resolution. 

\section{Conclusion}
\label{sec:con}

We have performed modulation transfer spectroscopy on a hot tellurium cell using an in-house ECDL laser at $455.4~$nm constructed from a commercially available Nichia diode. The laser is built for trapped barium ion experiment. For the first time we have observed seven resonance lines in the neighbourhood of barium ion dipole transition $S_{1/2}-P_{3/2}$, the closest being only $79~$MHz away. All the observed transitions are having signal-to-noise ratio of more than $10$, except for line no.$~7$ where the ratio is around $5$. The resonance line no.$~5$ closest to the barium transition has a S/N of more than $80$ leading to a robust frequency lock. These measurements will allow new reference for precision barium ion experiment which varies from fundamental physics to quantum information processing. Moreover the barium ion  $S_{1/2}-P_{3/2}$ transition is also used for Raman pumping into the dark $D_{5/2}$ state where absolute locking reference for the $455.4~$nm laser will make the experiments more stable and robust against frequency drifts. In addition, Te$_2$ is already an established reference for the other barium transition namely the $S_{1/2}-P_{1/2}$. Therefore, we believe that our newly measured references will further advance the toolbox of barium ion precision experiments.

\section{Acknowledgement}

DDM would like to acknowledge the contribution of Dr. Riadh Rebhi in developing a part of the setup used for performing this experiment. TD would like to acknowledge the contribution of Noah Van Horne in building some of the electronics.

\section{Funding Information}

This research is supported by the National Research Foundation Singapore under its Competitive Research Programme (CRP Award No. NRF-CRP14-2014-02)

\bigskip




\end{document}